\documentclass[manuscript]{aastex631}
\usepackage{amsmath}

\begin{document}
\title{Analysis and Modeling of High-Frequency Emission and Deep Seismic Sources of Sunquakes}

\correspondingauthor{John T. Stefan}
\email{jts25@njit.edu}

\author[0000-0002-5519-8291]{John T. Stefan}
\affiliation{Department of Physics, New Jersey Institute of Technology, Newark, NJ 07102}

\author[0000-0003-0364-4883]{Alexander G. Kosovichev}
\affiliation{Department of Physics, New Jersey Institute of Technology, Newark, NJ 07102}
\affiliation{NASA Ames Research Center, Moffett Field, Mountain View, CA 94035}

\begin{abstract}
Recent work published by Lindsey et al find evidence for a deep and compact seismic source for the sunquake associated with the 2011 July 30 M9.3 flare, as well as seismic emission extending up to 10 mHz. We examine the sunquake independently, and a possible wavefront is found in the 8 mHz band, though no wavefront is easily discernible in the 10 mHz band. Additionally, we perform numerical simulations of seismic excitation modeled with the reported parameters and changes in the power spectra with increasing depth of the excitation source are examined. It is found that the peak frequency decreases for increasing depths, but a shallow minimum is indicated between $z=0$ and $z=-840$ km. Analysis of the suspected wavefront of the M9.3 sunquake finds that the power spectrum of the reported seismic emission is close to that of background oscillations, though with a peak frequency noticeably lower than the background peak. Additionally, it is found that the amplitude of the source estimated by Lindsey et al (2020) is too low to produce the observed wavefront.
\end{abstract}

\section{Introduction}

Since the initial discovery of sunquakes by \cite{sq_disc}, much work has been done to determine the mechanisms behind the seismic emission as well as the development of new helioseismic tools for their study. In particular, helioseismic holography \citep{hh} has become a popular technique and was recently employed in the analysis of seismic emission associated with the 2011 July 30 M9.3 flare \citep{Lindsey}. The authors use the focus-defocus method, which---in short---adjusts the depth of the holographic focal plane until a particular feature reaches maximal sharpness, in order to determine the depth of the feature \citep{fdf}. The focus-defocus method is useful not only because it can identify the source depth of seismic emission, but can also determine the amplitude within a given frequency band. Using this method, \cite{Lindsey} find evidence for deep, high-frequency seismic emission. Here, we independently examine this event and compare its spectral properties with those of a confirmed sunquake, as well as simulations of seismic emission at increasingly large depths of the seismic source.

\section{Data and Methods}

\begin{figure}
    \centering
    \includegraphics[width=0.5\linewidth]{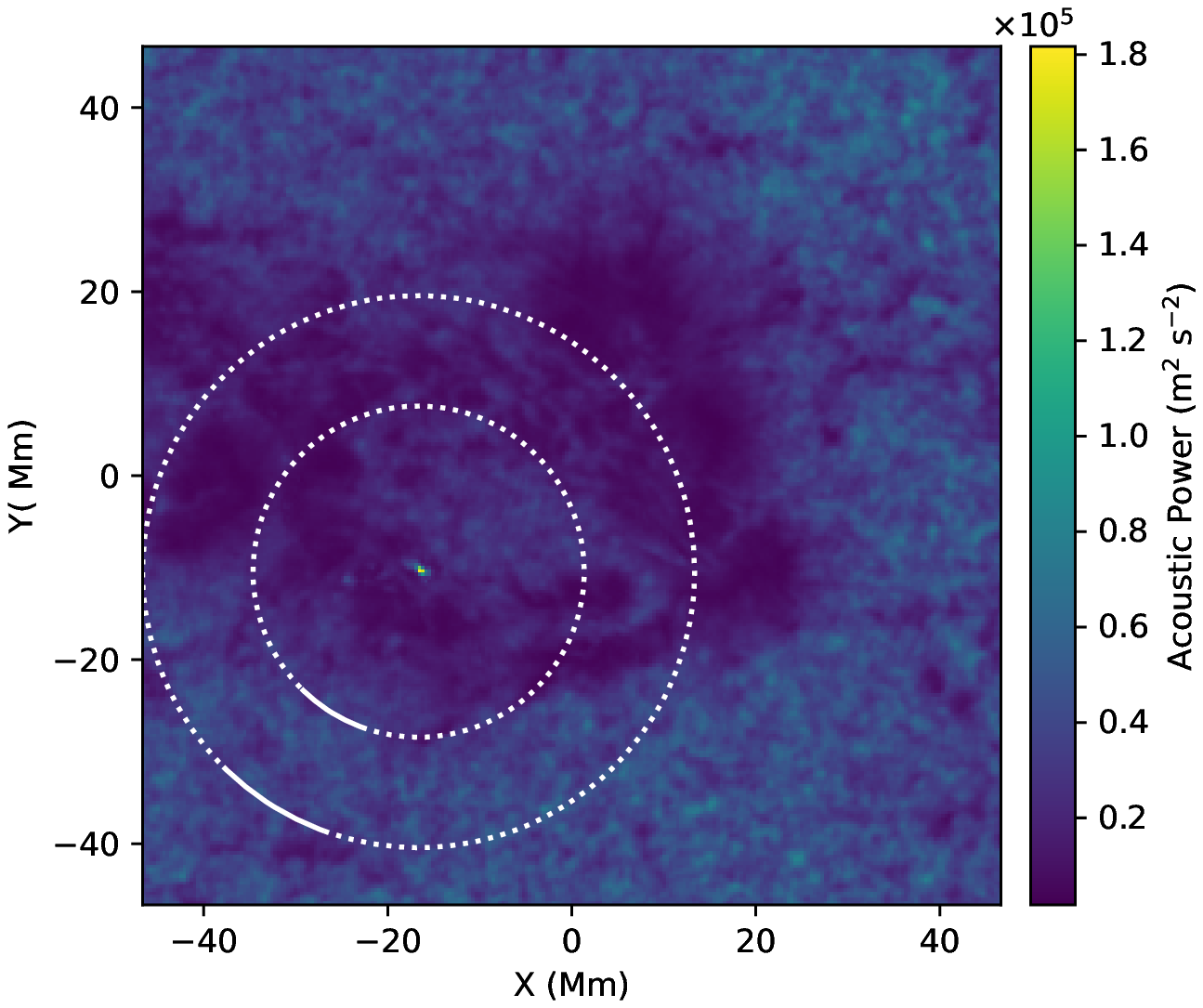}
    \caption{Acoustic power map of the area of interest. The seismic source location can be seen at the center of the concentric dotted rings, where the acoustic power is greatly enhanced. The extent of averaging for the time-distance diagram is displayed as solid white lines, with the inner arc placed at a distance of 18 Mm and the outer arc at 30 Mm. The dotted circles are added as a visual guide.}
    \label{fig:area}
\end{figure}

We obtain the tracked Dopplergram in the region of the sunquake, centered at heliographic latitude $+16^{\circ}$ and Carrington longitude $-31.2^{\circ}$, spanning two hours beginning at 02:00:00 UTC with 45 second cadence. The Dopplergram is remapped to a Postel's projection with resolution 0.06 deg pixel$^{-1}$ and total spatial extent $15.36^{\circ}$ by $15.36^{\circ}$. Finally, the time-difference is taken to increase the contrast of rapidly changing features, namely the sunquake wavefront. We locate the sunquake source by identifying the maximum of the time-differenced velocity, and the start time is visually adjusted to coincide with the initial excitation. Finally, we average the signal in {successive, 25$^{\circ}$} arcs about the excitation location for each frame in the Dopplergram to produce the initial time-distance diagram. {The averaging extent is shown by the solid white arcs in the acoustic power map of the area of interest, Figure \ref{fig:area}; the direction of averaging is chosen so that minimal damping of the wavefront occurs in areas of strong magnetic field. The areas where the magnetic field is strong are visible in the power map as regions of depressed acoustic power.}

As a comparison, we model a sunquake with the excitation parameters derived from the results of \cite{Lindsey}. The authors directly state an excitation depth of z=$-1150\pm120$ km, and the remaining parameters are estimated from the authors' focus-defocus depth diagnostics. We estimate the FWHM of the planar source function to be $1600$ km and the magnitude is derived from the 10 mHz source density, approximately $25$ m s$^{-1}$. The 3D acoustic model we use is linear with respect to the source amplitude \citep{Stefan}, and the resulting wavefront's amplitude can be scaled to remove any discrepancies with observations. We consider the transfer of momentum to be instantaneous such that the time-dependence of the source is a delta function at $t=0$.

As described in \cite{Stefan}, the model is semi-spectral, with the radial and time derivatives evaluated with finite differences and the horizontal derivatives evaluated spectrally. The Standard Solar Model \citep{ssmodel} is used as a background mesh, with some modification of the buoyancy frequency so that it is zero where the background mesh would prescribe it as negative, to remove the convective instability. The lower boundary is fixed at $r=0$, and the upper boundary allows outgoing oscillations to pass through without inducing any unphysical reflections.

In addition to modeling the excitation with the previously mentioned parameters, we also model excitations in increments of $420$ km to a maximum depth of $2100$ km. These additional simulations use the same 25 m s$^{-1}$ amplitude and horizontal FWHM of $1600$ km as the previously described z=$-1150$ km model, and similarly treated as an instantaneous transfer of momentum. The power spectral density is then computed from the corresponding time-distance diagrams, evaluated at horizontal distances where the potential wavefront of the observed sunquake is visible. The power spectral density is also similarly computed for a quiet sun region, in order to compare the frequency content and dependence of the modeled sunquakes.

\section{Results}

\begin{figure}
    \centering
    \includegraphics[width=\linewidth]{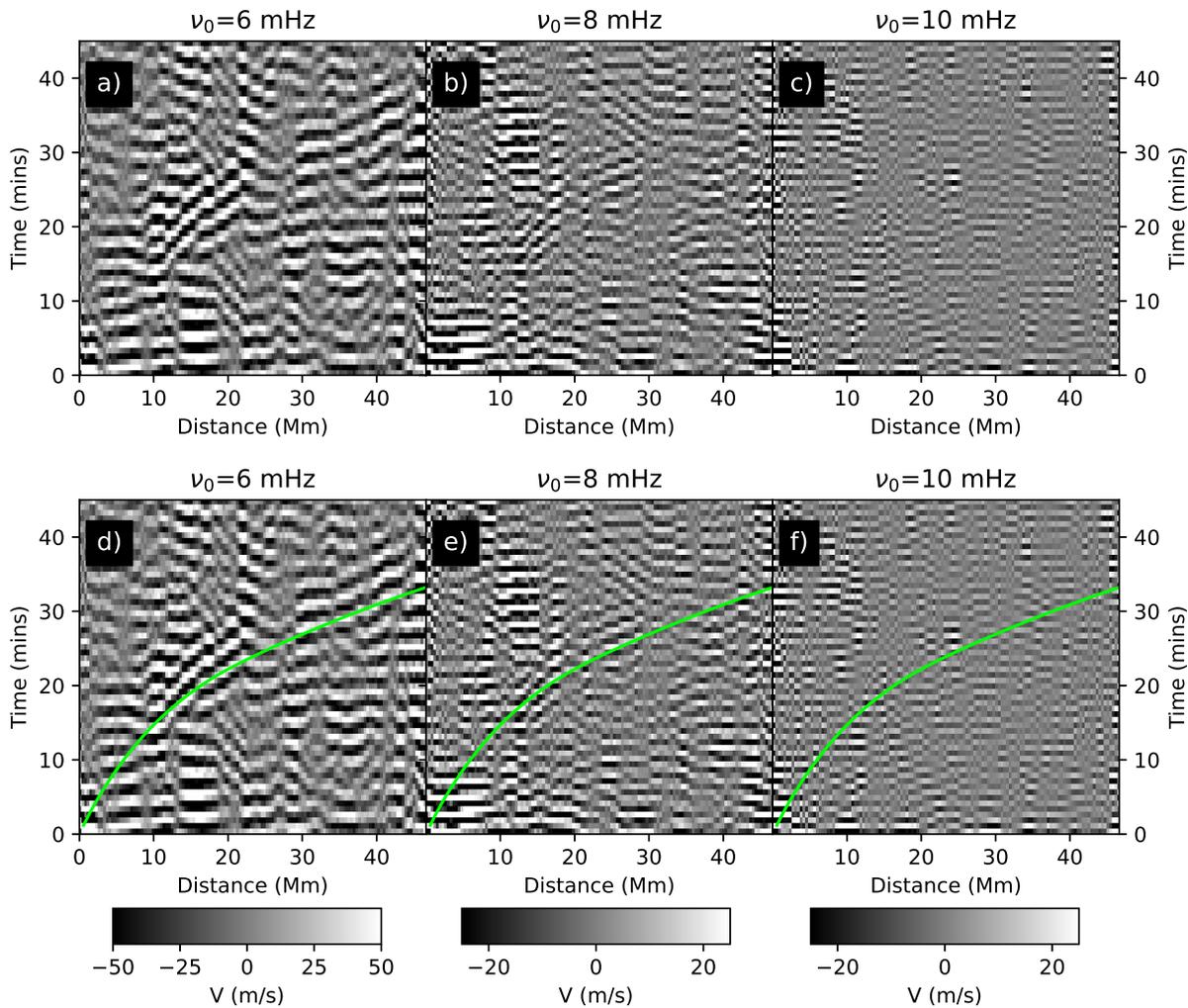}
    \caption{Time-distance diagrams for the seismic event associated with the M9.3 Flare, filtered around 6 mHz (a,d), 8 mHz (b,e), and 10 mHz (c,f). {The bottom row contains the same time-distance diagrams, with the time-distance curve predicted by asymptotic ray theory overlaid in green.}}
    \label{fig:td}
\end{figure}

We first examine the observational time-distance diagram for the reported sunquake obtained from HMI Dopplergrams, which have been treated with a flat-top Gaussian frequency filter to isolate oscillations centered at 6, 8, and 10 mHz. For these filters, oscillation power is untreated within $0.5$ mHz of the central frequency, and a Gaussian cut-off beyond $0.5$ mHz of the central frequency with $\sigma=0.5$ mHz. These time-distance diagrams are displayed in Figure \ref{fig:td} where the theoretical time-distance relation in the asymptotic ray approximation \citep{Gough} is overlaid in green in the second row. {The theoretical curve appropriately describes the shape of the time-distance relation, but may precede the center of the observed wavefront as the theoretical arrival time does not necessarily coincide with the group travel time.} The wavefront is distinctly visible in the 6 mHz band between 10 and 20 Mm, and again beginning at 30 Mm to the edge of the observational domain. In the 8 mHz band, we see a possible wavefront between horizontal distances of 10 and 30 Mm just above the theoretical time-distance curve. The wavefront is not well-resolved in the 10 mHz band, but some features indicative of a wavefront can be seen below the theoretical time-distance curve between distances of 28 and 35 Mm. In both cases, the wavefront is admittedly weak with amplitude on the order of background oscillations. Furthermore, the $6$ mHz band time-distance diagram shows a clear wavefront, well-aligned with theoretical curve from 10 Mm onward.

{ In the $6$ mHz time-distance diagram, the maximum wavefront amplitude between 10 and 20 Mm is 44 m s$^{-1}$ compared to the background maximum amplitude of 37 m s$^{-1}$, with a signal-to-noise ratio of $1.1$. For the $8$ mHz time-distance diagram in the same distance range, the maximum wavefront amplitude is $44$ m s$^{-1}$ compared to the background $37$ m s$^{-1}$, with a S/N of $1.19$. Finally, the suspected wavefront in the $10$ mHz time-distance diagram has a maximum amplitude of $11$ m s$^{-1}$ equal to the background maximum, with a S/N of $1$. In general, there will be less acoustic power in higher frequencies, so the decreases in both wavefront and background amplitudes are expected. The slight increase in S/N at $8$ mHz, compared to $6$ mHz, is also expected as this is significantly above the photospheric acoustic cut-off frequency, and acoustic power in this frequency domain comes predominantly from impulsive sources.}

We next compare the reported M9.3 sunquake's power spectrum with the quiet Sun spectrum and the spectrum of the visually confirmed sunquake associated with the X9.3 flare of 2017 September 6 \citep{Stefan}, {which are all computed from the corresponding time-distance diagram in the 1 Mm region about a horizontal distance of 18 Mm. For the quiet Sun, we first perform a 3-dimensional Fourier transform and azimuthally average the x- and y-wavevectors, resulting in a typical k-$\omega$ diagram. An inverse Fourier transform returns the data in the form of a time-distance diagram, essentially the quiet Sun's Green's function.} The power spectra are normalized, {with respect to the corresponding curves maximum spectral power}, as the oscillations for each case do not have similar amplitudes, and we are particularly interested in how the relative power is distributed between frequencies. The power spectra of both sunquakes is computed in the same range of horizontal distances between 10 and 20 Mm. We see in Figure \ref{fig:back_comp}a that the distribution of power for the reported M9.3 sunquake is very close to the quiet Sun background, though the spectrum peaks at a slightly lower frequency than the background. Additionally, we see no evidence for high-frequency oscillations at these distances, despite the power spectrum of the X9.3 sunquake (Figure \ref{fig:back_comp}b) showing enhancement in the high frequencies up to 7 mHz. The power spectrum for the M9.3 sunquake is superficially similar to the X9.3, with a sharp increase in power at the peak frequency and a somewhat {exponential} decrease as the frequency increases.

\begin{figure}
    \centering
    \includegraphics[width=0.75\linewidth]{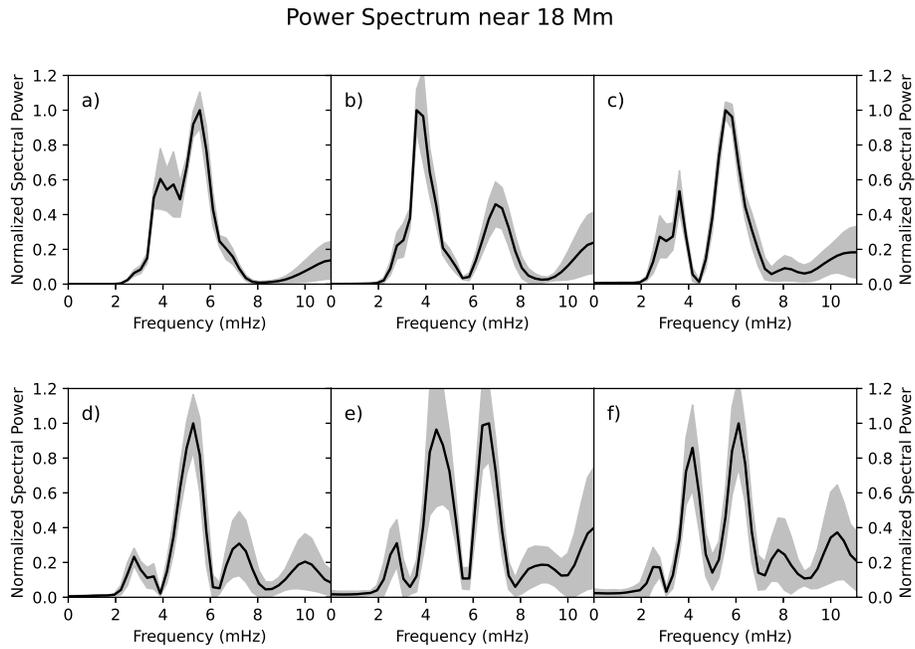}
    \caption{Comparison between the power spectral density of surface oscillations for sunquakes excited at $z=0$ km (a), $z=-420$ km (b), $z=-840$ km (c), $z=-1260$ km (d), $z=-1680$ km (e), and $z=-2100$ km (f).}
    \label{fig:freq_comp}
\end{figure}

Finally, we look at the oscillation power spectra of the simulated sunquakes with excitation depths increasing in increments of 420 km, to determine what spectral properties can be expected of deep seismic emission. Figure \ref{fig:freq_comp} shows the normalized power spectra computed for horizontal distances in the 1 Mm neighborhood of 18 Mm horizontal distance. {For all but the $z=-420$ case (Figure \ref{fig:freq_comp}b), which shows significant power around $4$ mHz, the peak frequency is at $5.5$ mHz or higher. While each simulated spectrum contains multiple peaks, the power spectra below excitation depths of $z=-1680$ km develop two peaks of nearly equal magnitude.} There is additional high-frequency ($8+$ mHz) content in all of the simulated cases, with the relative amplitude increasing with increasing depth {at $z=-1680$ km and below}, which is found in neither the X9.3 nor M9.3 sunquakes' power spectra. {Making a qualitative comparison, it appears that the power spectrum of the M9.3 sunquake is most similar to the $z=-420$ km and $z=-840$ km simulations, though the observed sunquake lacks a peak in the 6-8 mHz range. The X9.3 sunquake is most similar to the $z=-1680$ km simulation, with two distinct peaks between 4 and 6 mHz, and between 6 and 8 mHz.}

\section{Summary}

\begin{figure}
    \centering
    \includegraphics[width=0.75\linewidth]{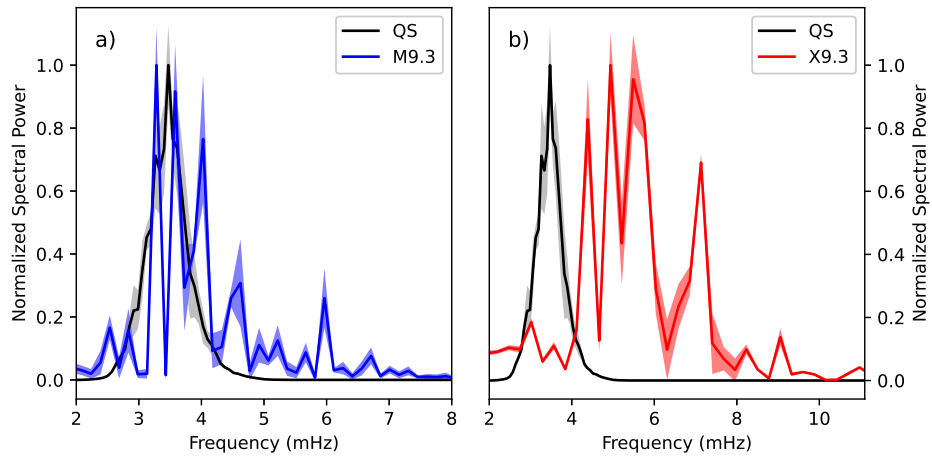}
    \caption{Comparison between the power spectral density of surface oscillations for the reported sunquake (a, blue) and the confirmed sunquake associated with the X9.3 flare of 2017 September 6 (b, red). The power spectrum of background oscillations is displayed in black.}
    \label{fig:back_comp}
\end{figure}

From the perspective of solar oscillation theory, we should expect a sunquake's peak frequency to decrease with increasing depth; in general, p-mode eigenfunctions with a deeper maximum have correspondingly lower frequencies. This is certainly true for {two deepest} test cases at $z=-1680$ km {and $z=-2100$ km}, though the transition from $z=-420$ km to $z=-840$ km actually shows an increase in the peak frequency. {A similar transition occurs from $z=-1260$ km to $z=-1680$ km, though this transition is less clear considering the relative magnitude of the two peaks in the $z=-1680$ km power spectrum. It is possible that a strictly decreasing trend in the peak frequency continues below $z=-2100$ km, though additional simulations will need to be run in order to determine this.} In all of the test cases for $z<0$ km, however, the power spectrum peaks at multiple frequencies, and the separation is uniform below $z=-840$ km. We see some evidence of a secondary peak between $8$ and $10$ mHz for the X9.3 sunquake (Figure \ref{fig:back_comp}), though there are no additional peaks for the M9.3 sunquake. While, in general, the high-frequency emission can be detected in sunquakes, it is substantially weaker in power than the primary emission at $5-6$ mHz.

Furthermore, our simulations for deep seismic emission, with maximum amplitude of 25 m s$^{-1}$, produce wavefront amplitudes on the order of 10 m s$^{-1}$, an order of magnitude lower than in unfiltered observations of the M9.3 sunquake. While we cannot rule out the depth of seismic emission reported by \cite{Lindsey}, the extremely low peak frequency can be indicative of shallow excitation or far deeper than is considered in this work. Additionally, we find that the amplitude estimate for the deep seismic source is too weak to produce the observed wavefront. {In order to reproduce the wavefront amplitude, the initial excitation itself must also have greater amplitude. Alternatively, deeper excitation with the same source amplitude may increase the observed wavefront, as more momentum is deposited at the source location.}

{The authors acknowledge support of the NSF grant 1916509 and NASA grants NNX14AB68G and 80NSSC19K0630.}

\newpage

\end{document}